\documentclass[aip,preprint,superscriptaddress,a4paper]{revtex4-1}
\usepackage{graphics}
\begin{document}

\title{Setting new Cosmology constraints with ALMA}

\author{Hugo \surname{Messias}}
\affiliation{Departamento de Astronom\'ia, Faculdad de Ciencias F\'isicas y Matem\'aticas, Universidad de Concepci\'on, Chile}
\affiliation{Centro de Astronomia e Astrof\'isica da Universidade de Lisboa, Portugal}

\email{hmessias@oal.ul.pt}


\begin{abstract}
I make a short revision of Cosmology questions which ALMA was built to address. Without diving into much detail, I point out the ALMA specifications and strategies which are expected to provide a better handle of: the temperature evolution of the Cosmic Microwave Background (CMB) and the properties of its secondary anisotropies (such as the thermal and kinetic Sunyaev-Zel'dovich and the Ostriker-Vishniac effects); variability of dimensionless fundamental constants; H$_0$ and galaxy initial mass function by means of strong gravitational lensing; black hole science with the greatly expected Event Horizon Telescope.
\end{abstract}

\maketitle

\section{\label{sec:intro}Introduction}

In my (still short) experience as an extra-galactic Astronomer, I got used to associating the question ``What Cosmology are they using?'' to mere three parameters: vaccum or dark energy density ($\Omega_\Lambda$), baryonic plus dark matter density ($\Omega_{\rm M}$), and the Hubble ``constant'' (H$_0$). As we all know, this simplification hides a whole lot of information. With the thirst to know where do we come from, Mankind characterized the Universe as best as possible in order to understand it. It did so by discovering or creating \cite{Duff02a,Duff02b} parameters and equations, which enabled a better description of the Universe's behaviour, but not always its nature \cite{Alexeevich10}.

To be accepted, such theories were tested against observations \cite{Danjon46,Dyson20}. It goes without saying that the instrumentation available to Mankind played a key role in this quest for knowledge. While improved instrumentation and statistical work keep breaking degeneracies between models, theoreticians implement changes to their theories or propose new ones.

This manuscript deviates from my presentation's title at Cosmosur\,II, but not from its content. I focus on key subjects of Cosmology and how the Atacama Large (sub-)Millimetre Array (ALMA) may help us take a step forward towards understanding our Universe.

\section{\label{sec:firmm}Cosmology in the FIR-mm spectral regime}

\subsection{\label{sec:tcmb}Temperature evolution of the Cosmic Microwave Background}

The Cosmic Microwave Background (CMB) is by far the most famous evidence for the expansion of the Universe and its very hot young phase. It is regarded as a surface of last scattering, given it marks the time beyond which the hot ($\sim$3000\,K) Universe had expanded enough for the hydrogen to be able to combine with the free floating electrons. The resulting photons (with a perfectly blackbody-shaped energy distribution peaking in the optical) could then freely start their travel towards us.

The continuous expansion of the Universe redshifted this background radiation and it now peaks in the microwave spectral regime ($\sim$5\,cm), implying an observed temperature of $T_0=2.72548\pm0.00057$\,K \cite{Fixsen09}. The way the Universe expanded is a key parameter for the current observed temperature value and its evolution with cosmic time. From the standard model, one expects the Universe to expand adiabatically, such that $T(z)=T_0(1+z)^{1-\alpha}$, where $\alpha=0$ \cite{Lima00}. Values greater than zero (up to $\alpha=1$) would imply that matter creation happens and compensates for the effect of expansion (to the extreme where $T(z)=cte=T_0$).

The two techniques used to trace the evolution of $T_{\rm CMB}$ are that based on the spectrum of the Sunyaev-Zel'dovich (SZ) effect \cite{Zeldovich69,Sunyaev70} (as suggested by \citet{Fabbri78} and \citet{Rephaeli80}) and that based on the excitation level of line species observed in absorption against background quasars \cite{Noterdaeme11} assuming the intervenient medium is in thermodynamic equilibrium with the CMB radiation. The most recent compilations point to $\alpha$ being consistent with zero \cite{Muller13,Hurier14,Saro14}, hence adiabatic expansion.

The SZ effect has so far been limited to $z<1.35$ clusters \cite{Hurier14,Saro14}, and given their scarcity at higher redshifts and the requirement of a hot intra-cluster medium, one should turn to the observation of line species in absorption, extending this study to farther distances. Here is where ALMA will contribute the most, given it is at higher redshifts that smaller shifts in $\alpha$ imply the largest differences \cite[see for instance Figure\,1 in][]{Saro14}. The large collecting area (up to 5,650\,m$^ 2$ from the main full-ALMA array) in addition to the spectral capabilities of ALMA will allow for a large pointing survey towards bright high-redshift quasars with known foreground absorbers. The reader should also recall the natural positive FIR-mm $k$-correction observed for redshifted galaxies, which increases the number of bright background sources.

\subsection{\label{sec:acmb}Anisotropies in the Cosmic Microwave Background}

Another characteristic of the CMB is the existence of small temperature variations \cite[up to 100's of $\mu$K,][]{Smoot92,Bennett13,PlanckI} on top the large-scale dipole induced by the Earth's movement respective to the CMB frame\cite{Smoot77}. These are called anisotropies and are the relic of matter density variations in the primordial Universe, thus being an imprint of the physics regulating the Universe \cite[e.g.,][]{Bennett97}. These anisotropies are believed to be the seeds of galaxies up to the largest clusters we see today. For instance, a density fluctuation of around a degree, ten's of arcminutes, or a few arcminutes imply co-moving distances, respectively, of about 100's of Mpc (super-cluster seeds), 10's of Mpc (cluster seeds), or a few Mpc (galaxy seeds).

Density variations on the sub-degree scale (multipole number $l\sim100^{\rm o}/\theta\gtrsim100$) depend on the matter and energy density and curvature of the Universe, and have now been determined by all-sky surveys such as \textit{WMAP} \cite{Bennett13}, the Atacama Cosmology Telescope \cite{Swetz11} (ACT), the South Pole Telescope \cite{Carlstrom11} (SPT), and \textit{Planck} \cite{PlanckI}, up to $l\sim3000$\cite{Calabrese13,PlanckI}. Beyond these scales, features in the CMB power spectrum vanish exponentially due to finite thickness of the last scattering surface. But it is at these small scales that secondary anisotropies peak, such as the non-linear thermal and kinetic SZ effects \cite{Zeldovich69,Sunyaev70} (tSZ and kSZ) and the linear Ostriker-Vishniac (OV) effect\cite{Ostriker86,Vishniac87}. Both the SZ and OV effects are ways to trace structure evolution after the last scattering surface ($z<1100$) and thus provide a means to constrain cosmological parameters such as the state equation of dark energy ($w$), matter density ($\Omega_m$) and the \textit{rms} density fluctuation amplitude in spheres of radius 8\,h$^{-1}$\,Mpc ($\sigma_8$) with the SZ effect \cite[e.g.,][]{Reichardt13,Hasselfield13,PlanckXX}, or the epoch of reionization with the OV effect \cite{Jaffe98,Peebles98} (although its observability has been questioned by \citet{Scannapieco00}).

While ALMA is not a survey telescope, its sensitivity and long-wavelength coverage make this facility the ultimate follow-up tool by confirming clusters with lower masses than those presented by ACT, SPT and \textit{Planck} teams. This will be achieved either by a low-frequency continuum survey to detect the tSZ effect towards the candidate cluster and/or redshift confirmation of cluster members. Nevertheless, for ALMA to detect both the SZ and OV effects, the ``On-The-Fly'' fringe tracking observation mode \cite{Daddario00,Rodriguez12} will be key. Finally, the resolution and spectral coverage will make the distinction of SZ and OV anisotropies from primordial CMB anisotropies easier.

\subsection{\label{sec:natconst}Nature's Fundamental ``constants''}

As mentioned earlier, Man has tried to describe the Universe by means of equations and parameters with which to attempt understanding its nature. Some of these parameters regulate key laws of physics and are believed to be constant under the General Relativity assumption that laws of physics are the same everywhere in the Universe (justifying the name \textit{Fundamental ``constants''}). However, would these parameters be variable with cosmic time or space, our current view of Nature would need to be revised. Hence, the community has devoted significant effort trying to constrain the variation in two dimensionless parameters: the fine-structure constant ($\alpha=e^2 / 4\pi \epsilon_0 \hbar c$) and the proton-electron mass ratio ($\mu=m_p/m_e$). While $\alpha$ represents the strength of the interaction between electrons and photons, the $\mu$ is related to the ratio of the strong force to the electroweak scale \cite{Flambaum04}.

A lower-limit on the variations of $\alpha$ has been provided by studies in the Oklo uranium mine in Gabon (Africa) where 2\,Gyr ago ($z\sim0.15$) a natural fission reactor operated: $\delta\alpha/\alpha\leq4.5\times10^{-8}$~~\cite{Lamoreaux04}. These time scales led people to seek such variations in the spectra of distant bright quasars with known foreground absorbers. Variations of $\alpha$ or $\mu$ will induce shifts in the frequencies of the lines observed in absorption, with different species having different sensitivities to those variations. Much work has been done in the ultra-violet and optical ranges using this technique. A $\delta\alpha/\alpha$ dipole across the sky was found\cite{Webb11,King12} at a $4\sigma$ level. Regarding $\Delta\mu/\mu$, there are still contradicting results, but no more than a $2\sigma$ evidence for variability up to $z_{\rm abs}\sim2.7$\cite{Rahmani13,Rahmani14,Bagdonaite14}.

An alternative spectral range is the (sub-)mm. Here, there are multiple lines to target (special focus has been devoted to ammonia and methanol \cite{Flambaum07,Bagdonaite13}), providing a large statistical improvement, and background sources are expectedly brighter due to the positive $k$-correction. With its spectral and detection capabilities, ALMA will, together with atomic clock experiments \cite{Salomon04}, expectedly revolutionize this field. As an example, using Effelsberg-100m, IRAM-30m, and ALMA Cycle-0 data, \citet{Bagdonaite13} have found a null hypothesis on $\Delta\mu/\mu$ as low as 0.1\,ppm at $z_{\rm abs}=0.89$.

\subsection{\label{sec:gl}Strong Gravitational Lensing}

The study of galaxy-galaxy chance alignments resulting in gravitational lensing provides a direct view of the dark and baryonic mass distribution of the lens. As such, people have used them to try constrain the Initial Stellar Masses (IMFs) of the lens. So far, the results support a Salpeter IMF \cite{Auger10,Treu10,Barnabe13}. With ALMA, observations of the gas in the lenses will be enabled, thus providing gas masses and dynamics, allowing a complementary view to the stellar studies (even though some bias toward the gas-richer late-type lenses is expected).

ALMA will be the tool to study the background source \cite{Vieira13,Weiss13}. This is due to the already referred positive (sub-)mm $k$-correction together with the collecting area and $uv$-space sampling of ALMA. These factors allow short integration times and still obtain reliable high-resolution imaging of the background source. The latter will be most relevant to constrain the lens model, especially for source-inversion lens modelling methods \cite[e.g.,][]{Dye13}. On top of that, the spectral capabilities of ALMA allow for the study of the background source's dynamics and, eventually, improve further the lens modeling, providing a high signal-to-noise is achieved in different line-channels. Short integration times will also enable the assessment of time-delays ($\Delta t$) which depend on the size of the Universe, i.e., the Hubble constant (H$_0$). Although distinct values for H$_0$ are found using this technique, including significant deviations from the CMB estimates \cite{Kochanek06}, ALMA is expected to help unveiling the solution for such discrepancy.

Finally, I note that these systems make very good samples to address the points mentioned in Sections~\ref{sec:tcmb} and \ref{sec:natconst}, because some of the background light will cross the lens plane at a distance where the lens inter-stellar medium is still significant, thus imprinting the required absorption feature in the background source spectrum \cite{Bagdonaite13}.

\subsection{\label{sec:bheh}Black Hole Event Horizon}

One of the most impressive observations in the very near future, will be the observation of the black hole in the center of our own galaxy, alias Sgr\,A* \cite{Doeleman09}. With the whole suit of ALMA 12\,m antennas working together as a single element of a global 1.3\,mm Very Large Baseline Interferometer (VLBI), ALMA will work as a stable phase anchor. Together with at least some antennas in the North hemisphere, this VLBI network will enable the observation of the immediate surroundings of Sgr\,A* at high significance. Some of the key questions to be addressed by this project are: does general relativity hold in the strong field regime; will we observe an event horizon; can black hole spin be measured with spatially resolved orbits near the event horizon; and how do black holes accrete matter and create powerful jets?

The latter question is posed at the right time. Since 2002, Gillessen et al.\cite{Gillessen12} traced a gas cloud getting close to Sgr\,A*. Last year (2013), it reached maximum proximity and, at this very moment, gas accretion onto Sgr\,A* is likely happening. Although ALMA is still not ready to be included in the network, stations in Chile (such as APEX\cite{Gusten06}) will enable high resolution constraints to the observations. This is the time for the Event Horizon Telescope!

\section{\label{sec:almaspec}ALMA specifications}

For completeness, I redirect the reader to the ALMA website, where a document on current and full-ALMA specifications is available\footnote{Look in \textit{Documents and Tools} $>$ \textit{Early Science Primer} at: http://www.almascience.org/}. I should also mention that this document is updated every Cycle, as it is still a growing facility, planned to reach completion this year on the next. So keep tuned!

\begin{acknowledgments}

I thank the Cosmosur\,II SOC for organizing the conference and the talk opportunity.

\end{acknowledgments}

\bibliography{bibl}

\end{document}